\documentstyle[aps,preprint]{revtex}
\tighten
\begin{document}
\title{Analytic approximation and an improved method for computing 
  the stress-energy of quantized scalar
        fields in Robertson-Walker Spacetimes}
\author{Paul R. Anderson\cite{pra} and Wayne Eaker\cite{we}}
\address{Department of Physics\\Wake Forest University\\P.O. Box 7507\\
   Winston-Salem, NC  27109}
\maketitle
\begin{abstract}
An improved method is given for the computation of the stress-energy tensor
of a quantized scalar field using adiabatic regularization.  The 
method works for fields with arbitrary mass and curvature coupling in 
Robertson-Walker spacetimes and is particularly useful
for spacetimes with compact spatial sections.  For massless fields
it yields an analytic approximation for the stress-energy tensor 
that is similar in nature to those obtained previously for
massless fields in static spacetimes.
\end{abstract}

\vspace{.5in}

Adiabatic regularization is a very useful technique for numerical
computations of the stress-energy tensor for quantized fields
in cosmological spacetimes.  It has been developed for
Robertson-Walker, Bianchi Type I, and Gowdy $T^3$ spacetimes  
 ~\cite{P,PF,FP,FPH,Hu,Berger,B,AP} and has been used to numerically compute
the stress-energy for quantized scalar fields in these spacetimes  
 ~\cite{FPH,Berger,HP1,HP2,A34,SA,Berger1,HMM}.
It has been proven to be equivalent to point splitting in Robertson-Walker,
RW, spacetimes
 ~\cite{Birrell,AP}.

In this paper we present an improved method for computing
the stress-energy tensor of a quantized scalar field using adiabatic 
regularization.  The method works for fields with arbitrary mass and
curvature coupling in a general RW spacetime. 
It has two advantages over previous methods.  First it is more
easily applied to spacetimes with compact spatial sections such
as the case of a RW universe with positive spatial
curvature.  Second it results in an analytic approximation for
the stress-energy tensor which can be useful for massless fields
and fields with very small masses.  
The approximation is very similar in nature to
those obtained previously for massless fields in 
static spacetimes~\cite{Page,FZ,AHS}.  It
is different from the approximation obtained by summing over 
all terms in the DeWitt-Schwinger expansion
which contain factors of the scalar curvature~\cite{PT}.
By its nature the analytic approximation does not give much 
information about particle
production effects which are inherently nonlocal.  However,
it does give information about vacuum polarization effects and
can thus give good qualitative and quantitative
information about the behavior of the stress-energy tensor when
vacuum polarization effects dominate.  
A further advantage of the
analytic approximation is that when one is computing the
full renormalized stress-energy tensor for massless fields, the analytic
approximation allows one to separate out, at least to some extent,
 the vacuum polarization part from the particle production part.

In what follows we first discuss our method of computing
the stress-energy tensor for quantized scalar fields in RW spacetimes
and we derive the analytic approximation.  We next discuss the
validity and usefulness of the approximation in various cases and
finish by  comparing it to the full renormalized stress-energy
tensor in a particular case in which that tensor is known.  

To begin consider a free scalar field with arbitrary mass and curvature
coupling in a RW spacetime.  The metric for a general RW spacetime
can be written~\footnote{Throughout this paper we use units such that
 $\hbar = c = 1$.  The metric signature is $(+ - - -)$ and the 
 conventions for curvature tensors are $R^\alpha_{\beta\gamma\delta} = 
  \Gamma^\alpha_{\beta\gamma,\delta} - ...$ and $ R_{\mu\nu} = 
  R^\alpha_{\mu\alpha\nu}$.}
\begin{equation}
  ds^2 = a^2(\eta) \left(d\eta^2 - \frac{dr^2}{1 - K r^2} 
           - r^2 d \Omega^2 \right) \;\;\;.
\end{equation}
Here $K = 0, + 1, - 1$ correspond to the cases of zero, positive, and
negative spatial curvature respectively.  The field $\phi$ can be
expanded in the following manner~\cite{BDbook}
\begin{equation}
\phi(x) = \frac{1}{a(\eta)} \int d \tilde{\mu}(k) \left( a_k \, Y_k({\bf x}) \, \psi_k (\eta)
          + {a_k}^\dagger \, {Y_k}^*({\bf x}) \, {\psi^*}_k (\eta) \right) 
\end{equation}
with 
\begin{eqnarray}
  \int d \tilde{\mu}(k) &\equiv& \int d^3 k \;\;\; K = 0,  \nonumber \\
   &\equiv& {\int_0}^\infty d k \sum_{l,m}  \;\;\; K = -1 \nonumber \\
   &\equiv& \sum_{k,l,m} \;\;\; K = 1  \;\;\;. 
\end{eqnarray}
The spatial part of the mode functions $Y_k({\bf x})$ obeys the equation
\begin{equation}
\Delta^{(3)} Y_k({\bf x}) = - (k^2 - K) Y_k({\bf x}) \;\;\;.
\end{equation}
The time dependent part $\psi_k$ obeys the mode equation
\begin{equation}
{\psi_k}'' + (k^2 + m^2 a^2 + (\xi - 1/6) a^2 R) \psi_k = 0 \;\;.
\end{equation}
Here $m$ is the mass of the field and $\xi$ its coupling to the scalar
curvature $R$.  In spacetimes with the metric (1)
\begin{equation}
  R = 6 \left(\frac{a''}{a^3} + \frac{K}{a^2} \right) \;\;.
\end{equation}

The unrenormalized stress-energy tensor is given by the expressions\cite{B,AP}
\begin{mathletters}
\begin{eqnarray}
<0|{T_0}^0|0>_{u} &=& \frac{1}{4 \pi^2 a^4}\int d\mu(k) \left(|\psi_k'|^2 + (k^2+m^2 a^2) |\psi_k|^2 
             \right. \nonumber \\
      & & \left. \;\;\;\; + 6 \,\left(\xi - \frac{1}{6} \right) \left[\frac{a'}{a} (\psi_k {\psi^*_k}' 
    + \psi^*_k \psi_k') - \left(\frac{{a'}^{\,2}}{a^2} - K \right) |\psi_k|^2 \right] \right) \\
<0|T|0>_{u} &=& \frac{1}{2 \pi^2 a^4}\int d\mu(k) \left(m^2 a^2 |\psi_k|^2 
     + 6 \left(\xi - \frac{1}{6}\right) \left[|\psi_k'|^2 
      - \frac{a'}{a} (\psi_k {\psi^*_k}' + \psi^*_k \psi_k') \right. \right. \nonumber \\
  &  &  \left. \left.   - \left(k^2+m^2 a^2 + \frac{a''}{a} - \frac{a'^{\,2}}{a^2} 
           + \left(\xi-\frac{1}{6}\right) a^2 R \right) |\psi_k|^2 \right]\right)
\end{eqnarray}
\end{mathletters}
with 
\begin{eqnarray*}
 \int d \mu(k) &\equiv& \int_0^\infty dk\, k^2 \;\;\; K = 0, -1 \nonumber \\
               &\equiv& \sum_{k=1}^\infty k^2 \;\;\; K = 1 \;\;\;.
\end{eqnarray*}
To renormalize one subtracts off the renormalization counterterms
which come from a WKB expansion of the mode equation.  This is the usual method
of adiabatic regularization.  Schematically one has
\begin{equation}
  <T_{\mu\nu}>_{r} \,=\, <T_{\mu\nu}>_{u} - <T_{\mu\nu}>_{ad} \;\;\;.
   \label{eq:trenorm}
\end{equation}
These counterterms 
are given in Ref.\cite{B,AP}.  As discussed in detail in Ref.\cite{AP}
the adiabatic counterterms in the case $K = 1$ consist of an 
integral rather than
a sum over $k$.  The reason is that the counterterms should be local and
thus should be independent of whether the spatial sections are compact
or not.  This argument would also apply to K = 0, -1 RW spacetimes with
periodically identified spatial sections.  As a result there is an added 
difficulty in the case
of compact spatial sections in subtracting off the renormalization 
counterterms.

To improve on the method of adiabatic regularization we expand the
renormalization counterterms in inverse powers of $k$ keeping only 
terms which are ultraviolet divergent.  For the case of compact spatial
sections the integral is also changed into a sum.  We call the resulting expressions 
$<T_{\mu\nu}>_d$.  In a general RW spacetime they have the form
\begin{mathletters}
\begin{eqnarray}
<{T_0}^0>_{d} &=& \frac{1}{4 \pi^2 a^4}\int d\mu(k) 
               \left(k + \frac{1}{k} \left[\frac{m^2 a^2}{2} 
             - 3 \left(\xi - \frac{1}{6}\right) \left(\frac{a'^{\,2}}{a^2} 
             - K\right)\right] \right) \nonumber \\
                &  &  +\, \frac{1}{4 \pi^2 a^4} \int d\bar{\mu}(k) \frac{1}{k^3} 
                     \left(- \frac{m^4 a^4}{8} 
             - \frac{3 m^2 a^2}{2} \left(\frac{a'^{\,2}}{a^2} + K \right)
                    +  \left(\xi - \frac{1}{6}\right)^2\, 
                        {^{(1)}H_0}^0 \,\frac{a^4}{4}\right)  \\
<T>_{d} &=& \frac{1}{4 \pi^2 a^4} \int d \mu(k) \frac{1}{k} \,\left(m^2 a^2 - 
            6\left(\xi - \frac{1}{6}\right) \left(\frac{a''}{a}
            - \frac{a'^{\,2}}{a^2} \right) \right) \nonumber \\  
      &  & +\, \frac{1}{4 \pi^2 a^4} \int d \bar{\mu}(k) \frac{1}{k^3} \left(- \frac{m^4 a^4}{2} 
          - \left(\xi - \frac{1}{6}\right) 3 m^2 a^2 \left(\frac{a''}{a} 
      + K\right) \right. \nonumber \\
   &  & \left.  + \left(\xi - \frac{1}{6}\right)^2\, 
                   {^{(1)}H_\mu}^\mu \,\frac{a^4}{4} \right) 
\end{eqnarray}
\end{mathletters} 
with
\begin{eqnarray*}
  \int d\bar{\mu}(k) &\equiv& \int_\lambda^\infty dk k^2 \;\;\;K = 0, -1 
      \nonumber \\
                     &\equiv& \sum_{k = 1}^\infty k^2 \;\;\; K = 1 \;\;.
\end{eqnarray*}
Here $\lambda$ is an arbitrary lower limit cutoff and
\begin{mathletters}
\begin{equation}
 {^{(1)}H_{\mu\nu}} = 2 R_{;\mu\nu} - 2 g_{\mu\nu} \Box R - \frac{1}{2}
     g_{\mu\nu} R^2 + 2 R R_{\mu\nu}  \;\;\;.
\end{equation}
In a RW spacetime it has the components
\begin{eqnarray}
 {^{(1)}H_0}^0 &=& - \frac{36 a''' a'}{a^6} + \frac{72 a'' a'^{\,2}}{a^7}
  + \frac{18 a''^{\,2}}{a^6} + \frac{36 K a'^{\,2}}{a^6} - \frac{18 K^2}{a^4}
     \label{eq:h100} \\
 {^{(1)}H_\mu}^\mu &=& - \frac{36 a''''}{a^5} + \frac{144 a''' a'}{a^6}
  - \frac{216 a'' a'^{\,2}}{a^7} + \frac{108 a''^{\,2}}{a^6} 
   + \frac{72 K a''}{a^5} - \frac{72 K a'^{\,2}}{a^6} \;\;. \label{eq:h1tr}
\end{eqnarray}
\end{mathletters}
The renormalized stress-energy
tensor is then computed by subtracting and adding 
the quantity $<{T_\mu}^{\nu}>_{d}$ to Eq. (\ref{eq:trenorm}) with the result that
\begin{mathletters}
\begin{eqnarray}
 <T_{\mu\nu}>_{r} &\equiv& <T_{\mu\nu}>_{n} + <T_{\mu\nu}>_{an}   \\
 <T_{\mu\nu}>_{n} &\equiv& <T_{\mu\nu}>_{u} - <T_{\mu\nu}>_{d}   \\
 <T_{\mu\nu}>_{an} &\equiv& <T_{\mu\nu}>_{d} - <T_{\mu\nu}>_{ad} \;\;\;. 
\end{eqnarray}
\end{mathletters}
In general $<T_{\mu\nu}>_n$ must be computed numerically while 
  $<T_{\mu\nu}>_{an}$ can always be computed analytically.  The result is
\begin{mathletters}
\begin{eqnarray}
 <{T_0}^0>_{an} &=& \frac{1}{2880 \pi^2} \left(
  - \frac{1}{6}\, {^{(1)}{H_0}^0} + {^{(3)}{H_0}^0} - \frac{3 K(K-1)}{a^4}
        \right) + \frac{m^2}{288 \pi^2} {G_0}^0 \nonumber \\
   & & - \frac{m^2 K(K-1)}{192 \pi^2 a^2} 
         - \frac{m^4}{64 \pi^2} \left[\frac{1}{2} + \log\left(\frac{\mu^2 a^2} 
           {4 \lambda^2}\right) + \frac{1}{2} K(K+1)(2 C +
           \log \lambda^2 )\right] \nonumber \\
    & &       + \left(\xi - \frac{1}{6}\right)  
           \left[ \frac{{^{(1)}{H_0}^0}}{288 \pi^2} + 
            \frac{K(K-1)}{32 \pi^2 a^4} \left(1 + \frac{a'^{\,2}}{a^2} \right) \right. \nonumber \\
         &  &\left. + \frac{m^2}{16 \pi^2} {G_0}^0 \left(3 + \log\left(
           \frac{\mu^2 a^2}{4 \lambda^2}\right) + 
          \frac{1}{2} K(K+1) (2 C + \log \lambda^2 )\right) + 
          \frac{3 K m^2}{8 \pi^2 a^2} \right]  \nonumber \\
     &  &  + \left(\xi - \frac{1}{6}\right)^2  
               \left[ \frac{{^{(1)}{H_0}}^0}{32 \pi^2} \left(2 + 
          \log\left(\frac{\mu^2 a^2}{4 \lambda^2}\right) + 
          \frac{1}{2} K(K+1) (2 C + \log \lambda^2 )\right) 
         \right. \nonumber \\ &  & \left. - \frac{9}{4 \pi^2} \left(
    \frac{a'^{\,2} a''}{a^7} + \frac{K a'^{\,2}}{a^6} \right)\right] \label{eq:t00a}\\
 <T>_{an} &=& \frac{1}{2880 \pi^2} \left(
   -\frac{1}{6} \,{^{(1)}{H_\mu}^\mu} + {^{(3)}{H_\mu}^\mu} \right) 
    + \frac{m^2}{288 \pi^2} {G_\mu}^\mu 
    - \frac{m^2 K(K-1)}{96 \pi^2 a^2} \nonumber \\
   &  &    - \frac{m^4}{16 \pi^2} \left[1 + \log\left(\frac{\mu^2 a^2}
           {4 \lambda^2}\right) + \frac{1}{2} K(K+1)
           (2 C + \log \lambda^2)\right] \nonumber \\ 
   &  &    + \left(\xi - \frac{1}{6}\right)  
           \left[ \frac{{^{(1)}{H_\mu}^\mu}}{288 \pi^2} + 
           \frac{K(K-1)}{16 \pi^2 a^4} \left(\frac{a''}{a} - \frac{a'^{\,2}}{a^2} \right) 
    \right. \nonumber \\
    &  & \left. + \frac{m^2}{16 \pi^2} {G_\mu}^\mu 
           \left(3 + \log\left(\frac{\mu^2 a^2}{4 \lambda^2}\right)
         +\frac{1}{2} K(K+1)(2 C + \log \lambda^2) \right) \right. \nonumber \\  
 &  & \left. + \frac{3 K m^2}{8 \pi^2 a^2} - \frac{3 m^2 a'^{\,2}}{8 \pi^2 a^4} 
          \right] \nonumber \\
    &  &  + \left(\xi - \frac{1}{6}\right)^2 \left[ \frac{^{(1)}{H_\mu}^\mu}
         {32 \pi^2} \left(2 + \log\left(\frac{\mu^2 a^2}{4 \lambda^2}\right)
          + \frac{1}{2}K(K+1)(2 C + \log \lambda^2)\right) \right. \nonumber \\
  &  &  \left.  - \frac{9}{8 \pi^2} \left(
    \frac{4 a''' a'}{a^6} - \frac{10 a'' a'^{\,2}}{a^7} 
    + \frac{3 a''^{\,2}}{a^6} + \frac{4 K a''}{a^5} - \frac{6 K a'^{\,2}}{a^6} 
    +\frac{K^2}{a^4} \right)\right] \;\;.\label{eq:tra}
\end{eqnarray}
\end{mathletters}
Here $G_{\mu\nu}$ is the Einstein tensor with components
\begin{mathletters}
\begin{eqnarray}
 {G_0}^0 &=& - \frac{3 a'^{\,2}}{a^4} - \frac{3 K}{a^2} \\
 {G_\mu}^\mu &=& -\frac{6 a''}{a^3} - \frac{6 K}{a^2}
\end{eqnarray}
\end{mathletters}
and ${^{(3)}H_{\mu\nu}}$ is the tensor
\begin{mathletters}
\begin{equation}
 {^{(3)}H_{\mu\nu}} = {R_\mu}^\rho R_{\rho\nu} - \frac{2}{3} R R_{\mu\nu}
           - \frac{1}{2} R_{\rho\sigma} R^{\rho\sigma} g_{\mu\nu} 
           + \frac{1}{4} R^2 g_{\mu\nu} 
\end{equation}
with components
\begin{eqnarray}
{^{(3)}H_0}^0 &=& \frac{3 a'^{\,4}}{a^8} + \frac{6 K a'^{\,2}}{a^6}
                   + \frac{3 K^2}{a^4}  \\
{^{(3)}H_\mu}^\mu &=& \frac{12 a'' a'^{\,2}}{a^7} - \frac{12 a'^{\,4}}{a^8}
                 + \frac{12 K a''}{a^5} - \frac{12 K a'^{\,2}}{a^6} \;\;\;.
\end{eqnarray}
\end{mathletters}
For a massive field $\mu = m$ while for a massless field $\mu$ is an arbitrary
constant.  However, in the massless case the terms containing $\log(\mu^2)$ each have as
coefficients multiples of the tensor ${^{(1)}H_{\mu\nu}}$ which  comes 
from an $R^2$ term in the gravitational Lagrangian.  Thus the terms containing
$\log \mu^2$ simply correspond to a finite renormalization of the coefficient
of the $R^2$ term in the gravitational Lagrangian.  

Note that if $K = 1$ then $<T_{\mu\nu}>_{d}$ consists of a sum over
$k$ while $<T_{\mu\nu}>_{ad}$ consists of an integral over $k$ as previously
mentioned.  Thus either the integral must be converted to a sum or the sum to an
integral.  We have converted the sum to an integral using
the Plana sum formula~\cite{Pl1,Pl2,Pl3,Pl4}.  This formula is
\begin{equation}
 \sum_{n=m}^\infty f(n) = \frac{1}{2} f(m) + \int_m^\infty dx f(x) 
       + i \int_0^\infty \frac{dt}{e^{2 \pi t} - 1} [f(m+it) - f(m-it)] \;\;.
\end{equation}
Because of the way $<T_{\mu\nu}>_{d}$ is defined the third term in 
the Plana sum formula can be computed exactly.  In the traditional form
of adiabatic regularization one would  convert the integral
in the adiabatic counterterms to a sum using the Plana sum formula and
then substitute the result into Eq.(\ref{eq:trenorm}).  
However, if this is done then, for a massive
field, it is not possible to compute the third term in the Plana
sum formula analytically.  Thus the computation of the renormalized
stress-energy tensor is simplified somewhat by our method in
the $K=1$ case.   Clearly the same simplification would occur if
one was using compact spatial sections for $K = 0$ or $K = -1$ RW
spacetimes. 

By direct computation one finds that $<T_{\mu\nu}>_{n}$ and  $<T_{\mu\nu}>_{an}$
are separately conserved.  Thus $<T_{\mu\nu}>_{an}$ can be used as an
analytic approximation for the stress-energy tensor.  For $K = 0, -1$ 
spacetimes it does contain the arbitrary constant $\lambda$ so it is not
unique unless the coefficients of the log terms vanish.  However, as can
be seen by examining Eqs.(\ref{eq:t00a}) and (\ref{eq:tra}), changing the
value of $\lambda$ simply corresponds to a finite renormalization of the
cosmological constant, $R$, and $R^2$ terms in the gravitational Lagrangian.
It is important to note that this is only true when using  $<T_{\mu\nu}>_{an}$
as an analytic approximation.  The $\lambda$ dependent terms do not appear
in the full renormalized stress-energy tensor.  

Because it depends quartically and quadratically on the mass, $<T_{\mu\nu}>_{an}$
is not a good approximation in the large mass limit.  Previous numerical work~\cite{A34}
indicates that the relevant condition is likely to be $m a << 1$.  
The quantity $<T_{\mu\nu}>_{an}$ is also local in the sense that it depends on the 
scale factor and its derivatives at a given time $\eta$.  Thus it cannot
accurately describe particle production effects which are inherently nonlocal.
However when used as an analytic approximation, it has the potential to 
do a good job in describing vacuum
polarization effects.  For the case of the conformally invariant scalar
field the analytic approximation is exact.  For other cases, it
is not usually exact and may not always be quantitatively a good approximation,
but qualitatively it can still be very useful.  For example the renormalized
stress-energy tensor has been computed analytically by Bunch and 
  Davies~\cite{BunchDavies}
for a massless minimally
coupled scalar field in a $K=0$ universe undergoing a powerlaw expansion
of the form $a = \alpha t^c = \alpha^{1/(1-c)} (1-c)^{c/(1-c)} \eta^{c/(1-c)}$, 
with $dt = a d\eta$ the proper time.  They choose
what is effectively the ``out'' vacuum which means that there is no particle
production.  Letting $p = c/(1-c)$ we find that the Bunch Davies result can
be written in the form
\begin{mathletters}
\begin{eqnarray}
<{T_0}^0>_{r} &=& \frac{1}{2880 \pi^2} \left[ - \frac{1}{6} {^{(1)}H_0}^0 +
         {^{(3)}H_0}^0 \right] \nonumber \\
 & & + \frac{1}{1152 \pi^2} {^{(1)}H_0}^0 \left[\log\left(
     \frac{\mu^2 a^2 \eta^2}{6 p(p-1)} \right) - \psi(\frac{3}{2}+\nu) 
     - \psi(\frac{3}{2} - \nu) - \frac{4}{3} \right] \nonumber \\
 & & + \frac{1}{128 \pi^2 a^4 \eta^4}\, p(p-1)(p^2+3p+4) \\ 
<{T_\mu}^\mu>_{r} &=& \frac{1}{2880 \pi^2} \left[ - \frac{1}{6} {^{(1)}H_\mu}^\mu +
         {^{(3)}H_\mu}^\mu \right] \nonumber \\
 & & + \frac{1}{1152 \pi^2} {^{(1)}H_\mu}^\mu \left[\log\left(
     \frac{\mu^2 a^2 \eta^2}{6 p(p-1)} \right) - \psi(\frac{3}{2}+\nu) 
     - \psi(\frac{3}{2} - \nu) - \frac{4}{3} \right] \nonumber \\
 & & + \frac{1}{32 \pi^2 a^4 \eta^4}\, p(p-1)(3p^2+5p+4) \;\;\;.    
\end{eqnarray}
Here $\nu \equiv |1-3c|/(2 |1-c|)$.
The analytic approximation on the other hand gives (if we absorb the
infrared cutoff $\lambda$ into the arbitrary constant $\mu$)
\begin{eqnarray}
<{T_0}^0>_{an} &=& \frac{1}{2880 \pi^2} \left[ - \frac{1}{6} {^{(1)}H_0}^0 +
         {^{(3)}H_0}^0 \right] \nonumber \\
 & & + \frac{1}{1152 \pi^2} {^{(1)}H_0}^0 \left[\log\left(\frac{\mu^2 a^2}{4}\right) 
         + 2 \right] - \frac{1}{32 \pi^2 a^4 \eta^4}\, p (p-1) (3p^2 + p)  \\
 <{T_\mu}^\mu>_{an} &=& \frac{1}{2880 \pi^2} \left[ - \frac{1}{6} {^{(1)}H_\mu}^\mu +
         {^{(3)}H_\mu}^\mu \right] \nonumber \\
& & + \frac{1}{1152 \pi^2} {^{(1)}H_\mu}^\mu \left[\log\left(\frac{\mu^2 a^2}{4}\right) 
         + 2 \right] + \frac{1}{32 \pi^2 a^4 \eta^4}\, p (p-1) (3 p^2 + 15p + 4) \;.
\end{eqnarray}
\end{mathletters}
 From these expressions one sees that some of the terms in the analytic approximation
are identical to those in the exact expression.  Those that differ do so only by
coefficients which in most cases are of the same order of magnitude. 
Thus  $<T_{\mu\nu}>_{an}$  clearly can serve as a useful
approximation in this case.  Given this fact, one can immediately deduce for
example that if $\xi \ne 0 $ then the stress-energy tensor will continue to
have terms of the form
\begin{equation}
  \frac{1}{a^4 \eta^4} (c_1 + c_2 \log(\mu^2 a^2)) \;\;\;.
\end{equation}

We have presented an improved method to compute the stress-energy
tensor for a scalar field in a RW spacetime using adiabatic regularization.  
The method has a computational
advantage over the usual method for spacetimes with compact spatial sections
where the unrenormalized terms contain a mode sum and the adiabatic
counterterms an integral.  Using the method we have derived an analytic 
approximation for
the stress-energy tensor which is particularly useful for massless 
fields when vacuum polarization effects dominate.

\acknowledgements
  Part of this work was done  while P.\ R.\ A.\ was visiting Los Alamos
National Laboratory.  He would like to thank E. Mottola, S. Habib, and the rest
of the T8 group for their hospitality.  This work was supported in part
by Grant No.\ PHY-9800971 from the National Science Foundation.


\begin{references}

\bibitem[*]{pra}electronic mail address: anderson@wfu.edu
\bibitem[\dagger]{we}electronic mail address: eakerwm5@wfu.edu
\bibitem{P} L. Parker, Ph.D. thesis, Harvard University, 1966 (Xerox University 
Microfilms, Ann Arbor, Michigan, No. 73-31244), pp. 140-171 and App. CI.
\bibitem{PF} L.\ Parker and S.\ A.\ Fulling, Phys.\ Rev.\ D {\bf 9}, 341 (1974).
\bibitem{FP} S.\ A.\ Fulling and L.\ Parker, Ann.\ Phys.\ (N.\ Y.) {\bf 87}, 176 (1974).
\bibitem{FPH} S.\ A.\ Fulling, L.\ Parker, and B.\ L.\ Hu, Phys.\ Rev.\ D {\bf 10}, 3905 (1974).
\bibitem{Hu} B.\ L.\ Hu, Phys.\ Rev.\ D {\bf 18}, 4460 (1978).
\bibitem{Berger} B.\ K.\ Berger, Ann.\ Phys.\ (N.\ Y.) {\bf 156}, 155 (1984).
\bibitem{B} T.\ S.\ Bunch, J.\ Phys.\ A {\bf 13}, 1297 (1980).
\bibitem{AP} P.\ R.\ Anderson and L.\ Parker, Phys. Rev. D {\bf 36}, 2963 (1987).
\bibitem{HP1} B.\ L.\ Hu and L.\ Parker, Phys.\ Lett.\ {\bf 63A}, 217 (1977).
\bibitem{HP2} B.\ L.\ Hu and L.\ Parker, Phys.\ Rev.\ D {\bf 17}, 933 (1978).
\bibitem{A34} P.\ R.\ Anderson, Phys.\ Rev.\ D {\bf 32}, 1302 (1985); {\bf 33}
              1567 (1986).
\bibitem{SA} W.\ M.\ Suen and P.\ R.\ Anderson, Phys. Rev. D {\bf 35}, 2940 (1987).
\bibitem{Berger1} B.\ K.\ Berger, Phys.\ Lett.\ {\bf 108B}, 394 (1982).
\bibitem{HMM} S.\ Habib, C.\ Molina-Paris, and E.\ Mottola, manuscript in preparation.
\bibitem{Birrell} N.\ D.\ Birrell, Proc.\ R.\ Soc.\ London {\bf B361}, 513 (1978).
\bibitem{Page} D.\ N.\ Page, Phys.\ Rev.\ D {\bf 25 }, 1499 (1982).
\bibitem{FZ} V.\ P.\ Frolov and A.\ I.\ Zel'nikov, Phys.\ Rev.\ D {\bf 35}, 3031 (1987).
\bibitem{AHS} P.\ R.\ Anderson, W.\ A.\ Hiscock, and D.\ A.\  Samuel, Phys.\ Rev.\ Lett.\ {\bf 70},
   1739 (1993); Phys.\ Rev.\ D {\bf 51}, 4337 (1995).
\bibitem{PT} L.\ Parker and D.\ J.\ Toms, Phys.\ Rev.\ D {\bf 31}, 953 (1985); {\bf 31}, 2424 (1985).
\bibitem{BDbook} see for example N.\ D.\ Birrell and P.\ C.\ W.\ Davies, 
{\it Quantum Field Theory in Curved Space} (Cambridge University Press, Cambridge,
 England, 1982) and references contained therein.
\bibitem{Pl1} E.\ Lindelof, {\it Le calcul des Residues} (Gautier-Villars, Paris,
   1905).
\bibitem{Pl2} E.\ T.\ Whittaker and G.\ N.\ Watson, {\it A Course of Modern
  Analysis} (Cambridge University Press, London, 1927), p. 145, Exercise 7.
\bibitem{Pl3} G.\ Ghika and M.\ Visinescu, Nuovo Cimento {\bf 46A}, 25 (1978).
\bibitem{Pl4} T.\ C.\ Shen, B.\ L.\ Hu, and D.\ J.\ O'Conner, Phys.\ Rev.\
 D {\bf 31}, 2401 (1985).
\bibitem{BunchDavies} T.\ S.\ Bunch and P.\ C.\ W.\ Davies, J. Phys. A {\bf 11},
   13, (1978).
\end{references}
\end{document}